\documentclass[aps,prd,preprint,tightenlines,groupedaddress,showpacs]{revtex4}
\usepackage{epsf,epsfig,graphics,graphicx}

\begin{document}
%\preprint{MMC\#3}  
\title{Correlated Hybrid Fluctuations from Inflation with Thermal
    Dissipation}
\author{Wolung Lee}
\email{leewl@phys.sinica.edu.tw}
\affiliation{Institute of Physics, Academia Sinica, Taipei, Taiwan 11529,
Republic of China}
\author{Li-Zhi Fang}
\email{fanglz@physics.arizona.edu}
\affiliation{Department of Physics, University of Arizona,
         Tucson, AZ 85721}

\date{\today}

\begin{abstract}

We investigate the primordial scalar perturbations in the thermal dissipative inflation where the radiation component (thermal bath) persists and the 
density fluctuations 
are thermally originated. The perturbation generated in this model is hybrid, 
i.e. it consists of both adiabatic and isocurvature components. We calculate the 
fractional power ratio ($S$) and the correlation coefficient ($\cos\Delta$) 
between the adiabatic and the isocurvature perturbations at the commencing 
of the radiation regime. Since the adiabatic/isocurvature decomposition of 
hybrid perturbations generally is gauge-dependent at super-horizon scales 
when there is substantial energy exchange between the inflaton and the 
thermal bath, we carefully perform a proper decomposition of the perturbations.  
We find that the adiabatic and the isocurvature perturbations are correlated, 
even though the fluctuations of the radiation component is considered 
uncorrelated with that of the inflaton. We also show that both $S$ 
and $\cos \Delta$ depend mainly on the ratio between the dissipation coefficient 
$\Gamma$ and the Hubble parameter $H$ during inflation. The correlation is 
positive ($\cos\Delta > 0$) for strong dissipation cases where $\Gamma/H >0.2$, 
and is negative for weak dissipation instances where $\Gamma/H <0.2$. Moreover, 
$S$ and $\cos \Delta$ in this model are not independent of each other. The
predicted relation between $S$ and $\cos\Delta$ is consistent with the 
WMAP observation. Other testable predictions are also discussed.

\end{abstract}

\maketitle

%\narrowtext

\section{Introduction}

The recently released data of the Wilkinson Microwave Anisotropy Probe 
(WMAP) confirmed the earlier COBE-DMR's observation about the deficiency in 
fluctuation power at the largest angular scales~\cite{wmap,dmr4}. The amount 
of quadrupole and octopole modes of the CMB temperature fluctuations is 
anomalously low if compared to the prediction of the $\Lambda$CDM model. 
It implies that the initial density perturbations are significantly suppressed 
on scales equal to or larger than the Hubble radius. Models of 
structure formation with a cut-off power spectrum of perturbation on large 
scales provide a better fit to the CMB temperature fluctuations. The most 
likely cut-off wavelength derived from the WMAP data~\cite{bri} actually is 
the same as that determined by the COBE-DMR~\cite{fj,bfh}. 

The super-horizon suppression is difficult to make compatible with models 
which produce pure adiabatic (isentropic) perturbations. However, it might 
be explained if the perturbations are hybrid. The different behavior of 
adiabatic and isocurvature (entropic) perturbations around the horizon scale 
can be used to construct power spectra with a super-horizon suppression.
The WMAP data show not only a possible non-zero fraction of isocurvature 
fluctuations in the primordial density perturbations, also the correlation 
between the adiabatic and the isocurvature components~\cite{pei}. These 
results then turn into the constraints on the multi-component inflationary 
models, as the initial perturbations generated from these models are 
principally hybrid~\cite{kps}. The double and multi-field models have 
been extensively studied in this context~\cite{double}. 

In this paper we will investigate the hybrid perturbations created by an 
inflation with thermal dissipation, the warm inflation scenario~\cite{bf}. 
In the scheme of the thermal dissipative inflation the universe contains a 
scalar field and a thermal bath during the inflation era. The two components 
are coupled via the thermal dissipation. In addition to fitting the amplitude 
and the power law index of the power spectrum given by the COBE data~\cite{lf1}, 
the thermal dissipative inflation leads to a super-horizon suppression of the perturbations by a factor $\geq 0.5$~\cite{lf3}. Recently, it has been found that the warm inflation of a spontaneous symmetry breaking potential with strong dissipation 
is capable of accommodating a running spectral index $n$ of the primordial perturbations, and generally yields $n >1$ on large scales and  $n<1$ on 
small scales~\cite{hall}. Our purpose here is to study the fractional power 
of the isocurvature perturbations, as well as the cross correlation between 
the adiabatic and the isocurvature fluctuations in the thermal 
dissipative inflationary model.  

In contrast to a single or a double field inflations, the evolution of the 
universe in the thermal dissipative inflation does not need a stage of 
non-thermal post-inflationary reheating. As long as the damping coefficent 
$\Gamma$ satisfies the criterion given in ~\cite{lf1}, 
$\Gamma > (M/m_{\rm Pl})^4H $, where $m_{\rm Pl},\ M$, and $H$ stand for 
the Planck energy, the energy scale, and the Hubble expansion of the 
inflaton respectively, the dissipation is effective enough to make the 
temperature of the radiation component increase continuously during the 
inflationary epoch. The universe would eventually enter the radiation-dominated 
phase when the temperature is high enough so that the radiation component 
prevails. Since the evolution of entropy only depends upon the thermal 
dissipative process during inflation, the entropic perturbations are not 
contaminated by the entropy production in the reheating stage. Therefore, 
the primordial hybrid perturbations induced by the thermal dissipation can 
be calculated unambiguously.  

The dynamical background of the thermal dissipative inflation model has 
been investigated within the framework of quantum field theory. It has 
been shown that the dissipation may amount to the coupling of the inflaton 
to a large number of particle species~\cite{linde,ber3}. In this sense, the 
two-field model and the thermal dissipation model can be considered as two 
extremes among multi-component inflations. The former adds one more field 
to the single inflaton, while the later has a large number of additional 
fields. 
    
The adiabatic and the isocurvature perturbations in the thermal dissipative 
model have been estimated in \cite{lf2,tb}. Yet, these calculations are not 
immune from the problems induced by gauge issues which are crucial for 
thermal dissipative perturbations~\cite{lee}. In particular when interactions
between the inflaton and the thermal bath are substantial, the commonly used adiabatic/isocurvature decomposition is not gauge-independent on the ground 
of super-horizon. Therefore, we must take a full relativistic treatment to 
analyze the evolution of the hybrid perturbations generated in the 
thermal dissipative inflation. Moreover, the fluctuations of the radiation 
component have not been carefully considered in previous works. Although the 
energy fluctuations of the radiation component are always less than that of 
the inflaton field, they are not negligible in examining the relative phase 
between the adiabatic and the isocurvature perturbations.    

This paper is organized as follows. In \S II we introduce the thermal 
dissipative inflationary model in relativistic covariant form. The initial 
adiabatic-to-isocurvature ratio is given in \S III. Sec. IV presents a full relativistic calculation on the super-horizon evolution of adiabatic and 
isocurvature perturbations. The numerical result of the spectrum of the 
adiabatic-to-isocurvature ratio is also given in \S IV. We then summarize 
our findings in \S V. The appendices provide the necessary details of 
the relativistic theory of linear perturbations.

\section{The models of thermal dissipative inflation}

\subsection{The background field and radiation}

We consider a universe consisting of a scalar inflaton field $\phi$, and a 
radiation component which mimics the thermal bath. The total energy-momentum 
tensor is 
%eq1
\begin{equation}
T_{ab}=T_{(\phi)ab} + T_{(r)ab},
\end{equation} 
where subscripts $(\phi)$ and $(r)$ are respectively for the scalar field and 
the radiation, and the Latin indices run from 1 to 4.

The energy-momentum tensor can be decomposed into fluid quantities as~\cite{el}
%eq2
\begin{equation}
T_{ab}=\rho u_au_b+ph_{ab}+q_au_b+q_bu_a+\pi_{ab},
\end{equation}
in which $q_au^a=\pi_{ab}u^b=0$, $\pi_{ab}=\pi_{ba}$, and $u_a$ can be any 
timelike vector field, i.e. $u^au_a=-1$, which is generally taken to be the 
average velocity vector. The total energy density of matter measured by an 
observer $u^a$ is represented by $\rho$, while $p$ and $\pi_{ab}$ denote 
the isotropic and anisotropic pressures respectively. The quantity $q_a$ 
prescribes the energy flux relative to $u^a$. 

With $u_a$, the line element of the spacetime can be written as
%eq3
\begin{equation}
ds^2\equiv g_{ab}dx^adx^b=-\left(u_adx^a\right)^2+h_{ab}dx^adx^b,
\end{equation}
where $h_{ab}\equiv g_{ab}+u_au_b$ is a projection tensor which maps points 
into the rest space of the observer $u^a$. Hence at the space-time point 
$x^a$, the observer $u^a$ assigns to the event $x^a+dx^a$ a spatial 
separation $(h_{ab}dx^adx^b)^{1/2}$, and a time separation $u_adx^a$ 
from him. 

For a minimally coupled scalar field $\phi$, the Lagrangian is
%eq4
\begin{equation}
L=\frac{1}{2}\partial^{a}\phi\partial_{a}\phi -V(\phi),
\end{equation}
where $V(\phi)$ is a self-interaction potential. The fluid quantities of the 
energy-momentum tensor for the $\phi$ field are then
%eq5
\begin{eqnarray}
\rho_{(\phi)} = \frac{1}{2}(\phi,_a u^a)^2+V(\phi), & \ & \ \ \
p_{(\phi)} = \frac{1}{2}(\phi,_a u^a)^2-V(\phi), \\ \nonumber
q_{(\phi) a}=-\phi,_c u^c h^b_a\phi_{,b}, & \ & \ \ \
\pi_{(\phi)ab}=0.
\end{eqnarray}
Assuming the radiation component is a relativistic ideal fluid, we have 
%eq6
\begin{eqnarray}
\rho_{(r)}=\kappa T^4,  & \  &  \ \ \ p_{(r)}=(1/3)\rho_{(r)}, \ \ \\ 
\nonumber 
q_{(r)a}=0,   &  \  & \ \ \ \pi_{(r)ab}=0,
\end{eqnarray}
where $T$ is temperature, $\kappa=(\pi^{2}/30)g_{\rm eff}$ and $g_{\rm eff}$ 
is the effective number of degrees of freedom at temperature $T$. As 
mentioned previously, the thermal dissipative inflation is a multi-component 
model with very high multiplication, the parameter $g_{\rm eff}$ amounts to 
the actual number of components. For simplicity, we will absorb the factor 
$\kappa$ into $T$, and use $\rho_{(r)}=T^4$ in what follows. 

\subsection{Interactions between the scalar field and the thermal bath}

The total energy-momentum conservation of the system is governed by 
%eq7
\begin{equation}
T^{ab}_{~~;b}=0.
\end{equation}
The interactions between the $\phi$ field and the thermal bath are 
characterized by the force vectors defined as
%eq8
\begin{equation}
Q_{(\phi)a} \equiv T_{(\phi)a;b}^{~~b}, \hspace{6mm}
Q_{(r)a} \equiv T_{(r)a;b}^{~~b}, 
\end{equation}
Obviously, we have
%eq9
\begin{equation}
Q_{(\phi)a} + Q_{(r)a} =0.
\end{equation}

The interaction term $Q_{(i)a}$ ($i$ is for $\phi$ or $r$) can be further 
decomposed into 
%eq10
\begin{equation}
Q_{(i)a}\equiv Q_{(i)}u_{a}+J_{(i)a} \hspace{6mm} {\rm and} \hspace{6mm}
u^{a}J_{(i)a}=0.
\end{equation}
The quantities $Q_{(i)}$ and $J_{(i)a}$ are, respectively, the temporal and 
the spatial components of $Q_{(i)a}$. They describe the energy and the 
momentum exchange between the scalar field and the thermal bath. Substituting 
(10) into (8), one has
%eq11
\begin{equation}
Q_{(i)}= -u^aT_{(i)a;b}^b  \hspace{6mm} 
J_{(i)c}= h_c^aT_{(i)a;b}^b.
\end{equation}
 
The dissipation of the scalar field can be modeled by
%eq12
\begin{equation}
Q_{(\phi)a} = - Q_{(r)a}= -\Gamma (\phi_{,b}u^b)\phi_{,a},
\end{equation}
where $\Gamma$ can be a function of $\phi$, and is always positive. With 
Eq. (10) we have
%eq13
\begin{equation}
Q_{(\phi)}= - Q_{(r)}= -  \Gamma (\phi_{,a}u^a)^2,
\end{equation}
and
%eq14
\begin{equation}
J_{(\phi)c}  = -J_{(r)c}=  - \Gamma (\phi_{,b}u^b)\phi_{,a}h_{c}^a.
\end{equation}

\section{The Initial adiabatic and isocurvature perturbations}

\subsection{The background solutions}

To find the background solutions, we consider all quantities being uniform 
and isotropic. Accordingly, the space-time of the universe assumes the flat 
Friedman-Robertson-Walker metric,
%eq15
\begin{equation}
ds^2=\bar{g}_{ab}dx^adx^b= -dt^2+a^2\left[dr^2+
r^2\left(d\theta^2+\sin^2\theta d\phi^2\right)\right].
\end{equation}
The Einstein equations of the expanding universe yield
%eq16 
\begin{equation}
\dot{H} = -4\pi G\left(\dot{\phi}^2+\gamma\rho_{(r)}\right),
\end{equation}
and
%eq17
\begin{equation}
H^{2} = \frac{8\pi G}{3}\left[\rho_{(r)} +
\frac{1}{2}\dot\phi^{2} + V(\phi)\right ],
\end{equation}
where $\gamma=4/3$ represents the adiabatic index of the thermal radiation, 
$a$ denotes the cosmic scale factor, and $H=\dot{a}/a$ is the Hubble 
parameter. 

The equation of motion for the scalar field $\phi$ is
%eq18
\begin{equation}
\ddot{\phi} + 3H \dot{\phi} + \Gamma\dot{\phi} 
- a^{-2}\nabla^2\phi+ V_{,\phi}(\phi)=0.
\end{equation}
For the uniform background field $\phi$, the $\nabla^2\phi$ term in (19) can 
be ignored, and we have
%eq19
\begin{equation}
\ddot{\phi} + (3H + \Gamma) \dot{\phi} + V'(\phi)=0.
\end{equation}
where the $\prime$ denotes $\partial/\partial \phi$.

The equation of motion for the radiation component (the thermal bath) is 
derived from the first law of thermodynamics as
%eq20
\begin{equation}
\dot \rho_{(r)} + 3 H \gamma \rho_{(r)} = \Gamma\dot{\phi}^{2}.
\end{equation}
Thus, the quantity $\Gamma$ in (12) represents the dissipation ``coefficient", 
which describes the production of radiation from the $\phi$ field. 
Furthermore, for the background solution, Eq. (14) gives
%eq21
\begin{equation}
J_{(\phi)c}=J_{(r)c}=0.
\end{equation} 
This is expected as the uniform and isotropic scalar field comoves with the 
radiation field, the net momentum exchange between them vanishes.

When the potential energy $V$ is dominant, i.e. 
$V \gg \rho_{(r)} + \frac{1}{2}\dot\phi^{2}$, the Hubble parameter $H(t)$ 
depends largely on $V$, and the universe undergoes an inflation. In the 
slow-roll regime $V'' \ll 3H^2$, Eq. (19) yields
%eq22
\begin{equation}
\dot{\phi} \simeq - \frac{V'}{3H + \Gamma}.
\end{equation}
The scalar field approximates the trajectory 
%eq23
\begin{equation}
\phi = \phi_0 e^{\beta t}
\end{equation}
with $\beta \simeq -V''(0)/(3H+\Gamma) \ll H$. Thus, with Eqs. (19) and (20), the behavior of the radiation component at the inflationary phase can be characterized by
%eq24
\begin{equation}
T^4\equiv\rho_{(r)}\simeq \frac{\Gamma\beta^2}{4H}\phi^2 \simeq 
 \frac{\Gamma}{4H}\dot{\phi}^2.
\end{equation}
This solution actually can be obtained directly from (20) by considering the slow rolling condition $\dot \rho_{(r)} \simeq 0$~\cite{bf}. Hence, Eq. (24) is still true even when $\Gamma$ is $\phi$-dependent. Numerical solutions to Eqs. (19) and (20) show that (24) is indeed available when $\Gamma$ is a power law function of $\phi$~\cite{lf1}.  

Thus, the temperature of the thermal bath increases mostly during the 
inflation era. In the standard inflation scenario which corresponds to 
the $\Gamma=0$ case, the radiation is blown off as $\exp(-4Ht)$. Therefore, 
the solution [Eq. (24)] is independent of the initial value of $\rho_{(r)}$ 
when $t$ is greater than $1/H$. 

Equations (19) and (20) imply that the $\phi$ field dissipation would 
eventually heat up the universe, giving rise to the co-existence of the 
scalar field and a radiation component with temperature larger than the 
Hawking temperature, i.e.
%eq25
\begin{equation}
T > H.
\end{equation}
Under this condition, the large scale reheating is unnecessary for inflation
with dissipation. That is, the inflation regime will smoothly transfer to 
the radiation-dominated regime when $T$ is high enough, and the radiation 
component becomes dominant~\cite{bf}. Numerical solutions to (19) and (20) 
have shown the smooth transition~\cite{lf1}. This feature is critical for calculating the primordial entropic fluctuations because the initial 
perturbations will be unaffected by the large scale post-inflationary 
entropic process, such as the reheating.

\subsection{Initial perturbations of the scalar field}

The initial perturbations of the $\phi$ field is calculated by the linearly 
perturbed field equation upon the space-time background (15). Because the 
primordial fluctuations are produced well within the Hubble radius, one can 
use the calculation without considering the gravitational gauge~\cite{lf1}. 
For instance, the Fourier mode of the $\phi$ field perturbations with 
comoving wavenumber $k$ is characterized by a Langevin-like equation
%eq26
\begin{equation}
\frac{d \delta \phi}{dt}= - 
\frac{k^2a^{-2} + V^{''}(\phi)}{3H+\Gamma}\delta \phi + \eta.
\end{equation}
The noise term $\eta$ given by thermal fluctuations is Gaussian. The statistical property of the noise $\eta$ can be determined by means of the fluctuation-dissipation relation~\cite{bf}
%eq27
\begin{equation}  
\langle \eta \rangle=0,
\end{equation}
%eq28
\begin{equation}
\langle \eta({\bf k},t) \eta({\bf k'},t') \rangle
=D\delta_{\bf k,-k'}\delta(t-t'),
\end{equation}  
with $D=(3H^3T/2\pi)(3H+\Gamma)^{-1}$, and $\langle... \rangle$ denotes 
averaging over an ensemble. The relation that $D=(H^3/2\pi)$ for the quantum 
case without dissipation is easily recovered.

Taking the slow-roll condition $|V^{''}(\phi)| \ll 9H^2$ into account, the $V^{''}(\phi)$ term in Eq. (26) can be ignored at the horizon-crossing 
where $a/k = H^{-1}$. Accordingly, the correlation function of the 
fluctuations is given by
%eq29
\begin{equation}
\langle \delta \phi(t) \delta \phi(t') \rangle
\simeq D \frac{3H+\Gamma}{2H^2} e^{-(t-t')H^2/(3H +\Gamma)}, 
\ \ \ t > t',
\end{equation}
and the amplitude for the horizon sized perturbation $\delta\phi$ is about
%eq30
\begin{equation}
\delta \phi 
\simeq \left (\frac{3}{4\pi}HT  \right )^{1/2},
\end{equation}
which implies the thermal fluctuations dominate over the quantum ones in 
the era of $T > H$. Taking the ensemble average, we obtain the perturbation
$\delta\dot{\phi}$ just outside the horizon as
%eq31
\begin{equation}
\delta\dot{\phi} \simeq -
\frac{H^2}{3H+\Gamma} \delta{\phi}.
\end{equation}
The perturbations in the $\phi$ field energy density is then
%eq32
\begin{equation}
\delta\rho_{(\phi)}=\dot{\phi}\delta\dot{\phi}+V'\delta \phi 
  \simeq V'\delta \phi= - (3H+\Gamma)\dot{\phi}\delta \phi
  \simeq -\left[2(\Gamma_H+3)\Gamma_H^{-1/2}\right]\cdot HT^2\delta\phi,
\end{equation}   
where we have used the slow-roll condition (22), and $\Gamma_H\equiv\Gamma/H$.

\subsection{Initial perturbations of the thermal bath}

We consider the radiation component in thermal equilibrium with a temperature 
$T$. At sub-horizon scales, fluctuations in the thermal radiation can be 
estimated by $\delta\rho_{(r)}/\rho_{(r)}\simeq 1/\sqrt{n_r}$~\cite{landau}, 
where $n_r$ denotes the total number of the relativistic particles within 
the horizon $H^{-1}$. Since the photon number density is proportional to 
$T^3$, and the volume within the Hubble radius is about $(4\pi/3)H^{-3}$, 
we find that
%eq33
\begin{equation}
\frac{\delta\rho_{(r)}}{\rho_{(r)}} \simeq
 \sqrt{\frac{3}{4\pi}}\left(\frac{H}{T}\right)^{3/2}.
\end{equation}
Accordingly, the energy fluctuations caused by $\delta\rho_{(r)}$ are 
characterized by 
%eq34
\begin{equation}
\delta\rho_{(r)} \sim
 \frac{\delta\rho_{(r)}}{\rho_{(r)}}T^4\simeq \sqrt{\frac{3}{4\pi}HT}\cdot
  HT^2= HT^2\delta\phi,
\end{equation}
where Eq.(30) has been used. It should be emphasized that (34) qualifies 
only the relation between the variances of $\delta\rho_{(r)}$ and 
$\delta\phi$, but not their phases. Principally thermal fluctuations of 
radiation component is random-phased with respect to the $\phi$ field 
fluctuations, i.e. $\langle\delta\phi\delta\rho_{(r)}\rangle=0$. 

With the help of the background solutions (22) and (24), the relation 
between the fluctuations in the thermal energy and the $\phi$ field 
energy can be established,
%eq35
\begin{equation}
\delta\rho_{(r)} =\delta\rho_{(\phi)} \frac {HT^2}{V'(\phi)}=
   \frac {\Gamma_H^{1/2}}{2(\Gamma_H+3)}\delta\rho_{(\phi)}.
\end{equation}
Therefore, $\delta\rho_{(r)}$ is always less than $\delta\rho_{(\phi)}$ 
in situations either $\Gamma < H$ or  $\Gamma > H$. Apparently, the perturbed 
cosmic matter is just about isothermal. 

\subsection{The adiabatic vs. the isocurvature initial conditions}

The energy perturbations of the $\phi$ field and the radiation 
component can be decomposed into adiabatic (ad) and isocurvature (en) 
modes as
\begin{eqnarray}
%eq36
 \delta\rho_{(\phi)} &=&
   \delta\rho_{(\phi)}^{\rm ad} + \delta\rho_{(\phi)}^{\rm en},  \\
%eq37
\delta\rho_{(r)} &=& \delta\rho_{(r)}^{\rm ad} + 
    \delta\rho_{(r)}^{\rm en}.
\end{eqnarray}
By definition, the adiabatic perturbations $\delta^{\rm ad}$ are given by
%eq38
\begin{equation}
\delta^{\rm ad}=
 \frac {\delta\rho_{(\phi)}^{\rm ad} }{\rho_{(\phi)} +p_{(\phi)}}
=\frac{\delta\rho_{(r)}^{\rm ad}}{\rho_{(r)} + p_{(r)}},
\end{equation}
while the isocurvature mode satisfies
%eq39
\begin{equation}
\delta\rho_{(\phi)}^{\rm en} + \delta\rho_{(r)}^{\rm en} = 0.
\end{equation}
Therefore, the adiabatic perturbations can be rewritten as
%eq40
\begin{equation}
\delta^{\rm ad}=\frac{\delta\rho_{(\phi)}^{\rm ad} +
 \delta\rho_{(r)}^{\rm ad} }
{\rho_{(\phi)} +p_{(\phi)} + \rho_{(r)} + p_{(r)}}=
\frac{\delta\rho_{(\phi)} + \delta\rho_{(r)}}
{h_{(\phi)}+h_{(r)}},
\end{equation}
and the entropic perturbations are characterized by
%eq41
\begin{equation}
 \delta^{\rm en} =
\frac{\delta\rho_{(\phi)}^{\rm en}}{\rho_{(\phi)}+p_{(\phi)}}-
\frac{\delta\rho_{(r)}^{\rm en}}{\rho_{(r)}+p_{(r)}}=
\frac{\delta\rho_{(\phi)}}{h_{(\phi)}}-
\frac{\delta\rho_{(r)}}{h_{(r)}},
\end{equation}
where we have used $h_{(\phi)}\equiv \rho_{(\phi)}+p_{(\phi)} = \dot{\phi}^2$ 
from Eq.(5), and $h_{(r)}\equiv \rho_{(r)}+p_{(r)} = \Gamma_H\dot{\phi}^2/3$ 
from (6) and (24). The radiation part $\delta\rho_{(r)}/h_{(r)}$ in the 
above definition (41) is noteworthy as it may be compatible to the inflaton 
part $\delta\rho_{(\phi)}/h_{(\phi)}$, even though the condition 
$\delta\rho_{(r)} < \delta\rho_{(\phi)}$ is always fulfilled. That is, 
$\delta\rho_{(r)}$ must not be ignored when treating the isocurvature 
perturbation no matter how small it may be.
 
Considering the case when 
$\langle \delta\rho_{(\phi)}\delta\rho_{(r)} \rangle = 0$, one can obtain 
the variances for the adiabatic and the entropic perturbations by virtue of 
Eqs. (30), (32) and (35): 
%eq42
\begin{equation}
[\langle (\delta^{\rm ad})^2\rangle]^{1/2} =
\left(\frac{\rho_{(\phi)}}{h_{(\phi)}}\right)\cdot
\left[1+\frac{\Gamma_H}{4(\Gamma_H+3)^2}\right]^{1/2}
\left(1+\frac{\Gamma_H}{3}\right)^{-1},
\end{equation}
%eq43
\begin{equation}
[\langle (\delta^{\rm en})^2\rangle ]^{1/2} =
\left(\frac{\rho_{(\phi)}}{h_{(\phi)}}\right)\cdot
\left[1+\frac{9\Gamma_H^{-1}}{4\left(\Gamma_H+3\right)^2}\right]^{1/2}.
\end{equation}
The correlation between the two perturbation components is given by
%eq44
\begin{equation}
\langle \delta^{\rm en}\delta^{\rm ad}\rangle =
\left(\frac{\delta\rho_{(\phi)}}{h_{(\phi)}}\right)^2\cdot
\left[1-\frac{3}{4(\Gamma_H+3)^2}\right]\left(1+\frac{\Gamma_H}{3}\right)^{-1}.
\end{equation}
Equation (44) shows that even when the fluctuations of the $\phi$ field 
and the radiation are uncorrelated, the correlation between the adiabatic 
and the entropic perturbations can be significant. In these initial conditions 
all $\phi$ and $\dot{\phi}$-dependent quantities as well as $V'$, 
$h_{(\phi)}$, and $\Gamma_H$ are taken to be their values at the $k$-mode 
horizon-crossing times. 

\section{Super-horizon evolution of adiabatic and isocurvature modes}

\subsection{Adiabatic/isocurvature decompositions on super-horizon scale}

In super-horizon regions, physical quantities should be expressed in a 
gauge-independent fashion. However, the adiabatic/isocurvature 
decomposition of perturbations shown in the last section are not 
gauge-independent, i.e. either eqs.(40) or (41) are not gauge invariant 
variables. As usual~\cite{b88,h123}, we choose the condition to fix the 
coordinates and to make (40) and (41) physically meaningful. As shown in 
Appendix C, a gauge invariant (GI) variable $\zeta$ can be defined as 
%eq45
\begin{equation}
\zeta =
\varphi+ \frac{\delta\rho_{\rm tot}}{3(\rho_{\rm tot}+ p_{\rm tot})}=
\varphi+ \frac{\delta\rho_{(\phi)}+\delta\rho_{(r)}}
{3(h_{(\phi)}+h_{(r)})},
\end{equation}
where $\varphi$ is the perturbed variable of 3-space curvature [see Eq. (A2)].
 Thus, if we fix the coordinates by $\varphi=0$, the so-called 
uniform-curvature gauge (UCG) condition, then $\zeta$ becomes 
$\delta^{\rm ad}$ [Eq.(40)]. Thus, the initial condition and the evolution 
of the super-horizon-sized adiabatic perturbations can be described properly 
by the equations of $\zeta$ in the UCG.  

On the other hand, we can define another GI variable $\varpi$ using Eqs.(C11) 
and (C12) as
%eq46
\begin {equation}
{\varpi}= \frac{\delta\rho_{(\phi)}+ \delta J_{(\phi)}}{h_{(\phi)}} 
- \frac{\delta\rho_{(r)}+\delta J_{(r)}}{h_{(r)}},
\end{equation}
where $\delta J_{(\phi)}$ and $\delta J_{(r)}$ is the linear perturbed 
variables of energy exchange between the $\phi$ field and the thermal bath 
[Eqs.(14) and (A17)]. Evidently, $\varpi$ is nothing but $\delta^{\rm en}$ 
if replacing $\delta\rho_{(\phi)}$ and $\delta\rho_{(r)}$ in (41) respectively 
by $\delta\rho_{(\phi)} +\delta J_{(\phi)}$ and $\delta\rho_{(r)}+\delta J_{(r)}$. 
We may call $\varpi$ a modified isocurvature (entropic) perturbation. As 
a consequence, there exists no suitable coordinate fixing to make both 
$\delta^{\rm ad}$ and $\delta^{\rm en}$ simultaneously a GI variable as 
there is an energy exchange between the background components. In this 
regard, the decomposition of perturbations by (40) and (41) at super-horizon 
scales does not have clear physical meanings. Instead, one should decompose 
the super Hubble perturbations into the adiabatic and the modified isocurvature 
modes. The initial condition and the evolution of the modified entropic 
perturbation outside the horizon can then be described unambiguously by the 
equations of $\varpi$ in the UCG. Moreover, at the commencing of radiation 
regime when the inflation ends, $\dot\phi \simeq 0$, the energy ex
change in the background ceases and (46) reduces to (41). Therefore, 
perturbations defined by Eq. (46) can be explained as the isocurvature 
perturbations at the onset of the radiation-dominated epoch.

The initial condition of $\delta J_{(\phi)}$ can be estimated via (A17) as
%eq47
\begin{equation}
\delta J_{(\phi)} = - \delta J_{(r)} = \Gamma \dot{\phi} \delta\phi =  \left(2\Gamma_H^{1/2}\right)\cdot HT^2\delta\phi. 
\end{equation} 
Comparing to (30) and (32), $\delta J_{(\phi)}$ is always less than $\delta\rho_{(\phi)}$, but it can be larger than $\delta \rho_{(r)}$ if 
$\Gamma>H$. Since  $\delta J_{(\phi)} + \delta J_{(r)}=0$, we have 
$\delta\rho_{(\phi)} +\delta J_{(\phi)}+ \delta\rho_{(r)}+\delta J_{(r)}=
\delta\rho_{(\phi)} + \delta\rho_{(r)}$. Hence, (40) remain valid even 
when $\delta\rho_{(\phi)}$ and $\delta\rho_{(r)}$ are replaced by $\delta\rho_{(\phi)} +\delta J_{(\phi)}$ and $\delta\rho_{(r)}+\delta J_{(r)}$.

More importantly, since $\delta J_{(r)}$ involves the effect of $\dot{\phi}$, 
it is correlated with the $\phi$ field fluctuations. The perturbations 
$\delta J_{(r)}$ 
and $\delta \rho_{(\phi)}$ actually are in phase. Therefore, it is interesting 
to consider the case of imposing the phase correlation into the initial 
conditions. Connecting $\delta\rho_{(\phi)}$ and $\delta\rho_{(r)}$ by the 
relation (35), Eqs. (40) and (41) yield 
%eq48
\begin{equation}
\delta^{\rm ad} = \left(\frac{\delta\rho_{(\phi)}}{h_{(\phi)}}\right)\cdot
\left[1+\frac{\Gamma_H^{1/2}}{2(\Gamma_H+3)}\right] \left(1+\frac{\Gamma_H}{3}\right)^{-1},
\end{equation}
%eq49
\begin{equation}
\delta^{\rm en} = \left(\frac{\delta\rho_{(\phi)}}{h_{(\phi)}}\right)\cdot
\left[1-\frac{3\Gamma_H^{-1/2}}{2(\Gamma_H+3)}\right], 
\end{equation}
and
%eq50
\begin{equation}
\langle \delta^{\rm en}\delta^{\rm ad}\rangle =
\left(\frac{\delta\rho_{(\phi)}}{h_{(\phi)}}\right)^2\cdot
\left[1-\frac{3\Gamma_H^{-1/2}}{2(\Gamma_H+3)}\right]
\left[1+\frac{\Gamma_H^{1/2}}{2(\Gamma_H+3)}\right]
\left(1+\frac{\Gamma_H}{3}\right)^{-1}.
\end{equation}
Consequently, $\delta^{\rm en}$ and $\delta^{\rm ad}$ are in phase when 
$\Gamma_H > 0.217$, but are anti-correlated when $\Gamma_H < 0.217$. 

\subsection{Equations of the super-horizon evolution of perturbations}

The evolution of the linear perturbations outside the Hubble radius is 
described by
a set of equations of all perturbed matter variables, such as 
$\delta \rho_{(\phi)}$ and $\delta\rho_{(r)}$, as well as perturbed 
variables of the space-time metric. The sophisticated formalism of deriving 
these equations are given in Appendices. For the UCG, these equations are 
shown as (D1) - (D8) in Appendix D. With the 
solutions of Eqs. (D1) - (D8), it is straightforward to trace the 
evolution of relevant variables of the perturbations. 

The evolutionary features of the super-horizon-sized perturbations can 
be seen from Eqs. (D7) and (D8), which are
%eq51
\begin{eqnarray}
\delta\ddot{\phi} & +  &
  \left[(3H+\Gamma)-\frac{8\pi G}{3}
      \frac{\dot{\phi}^2}{H}\right]
  \delta\dot{\phi} + \left[\frac{k^2}{a^2}+V''+\frac{16\pi G}{3}
  \frac {\dot{\phi}V'}{H}\right]\delta\phi  \nonumber \\
& = &-\left[\frac{32\pi G}{3} V\frac{\dot{\phi}}{H} +
  \left(\Gamma\dot{\phi}+2V'\right)\right][\dot{\chi}+H\chi] 
  + \frac{2}{3}\frac{k^2}{a^2}\dot{\phi}\chi,
\end{eqnarray}
%eq52
\begin{eqnarray}
\ddot{\chi} & + & \left[H+\frac{32\pi G}{3}\frac{V}{H}\right]\dot{\chi} +
  \left[4\pi G\left(\frac{8}{3}V-\dot{\phi}^2-\frac{4}{3}\rho_{(r)}\right)- 
\frac{2}{3}\frac {k^2}{a^2}\right]\chi
\nonumber \\
& = & 
-4\pi G\left[-\frac{2}{3}\frac{\dot{\phi}}{H}\delta\dot{\phi}+
\frac{4}{3}\frac{V'}{H}\delta\phi\right].
\end{eqnarray}
These equations portray the linear evolution of the $k$-mode perturbation of 
the scalar field $\delta \phi$, and that of the shear $\chi$ of the space-time 
metric~\cite{b88}. 

The Friedmann Eq. (17) implies $(3H+\Gamma) \gg (8\pi G/3)(\dot{\phi}^2/H)$.
For modes at sub-horizon scales $(k^2/a^2) \geq H^2$, we have  
$|(k^2/a^2)+V''|\gg (16\pi G/3) (\dot{\phi}V'/H)]$. Accordingly, if taking 
the right hand side to be zero, Eq. (51) within the horizon is exactly the 
same as the linearized Eq. (18), or (26) with the slow-rolling condition. 
Since the initial perturbation of $\phi$ field is sub-horizon-scaled, and 
is governed by (18) or (26), it is consistently to assume the initial 
conditions for $\chi$ and $\dot{\chi}$ to be zero.  

Equations (51) and (52) can be regarded as two coupled oscillators with a 
time dependent mass, damping coefficients and coupling coefficients. Whenever 
the ``mass" becoming negative, the perturbations undergo a decaying or 
growing process. This gives rise to the gravitational clustering. The ``mass'' 
of the $\chi$-oscillator generally is positive for modes beyond the Hubble 
radius $(k^2/a^2) \ll H^2$. Therefore, the perturbations $\delta \phi$ will 
not be magnified by gravity outside the horizon, and will retain approximately 
their initial values. Consequently, variations in both $\dot{\delta}^{\rm ad}$ 
and $\dot{\delta}^{\rm en}$ are insignificant during their super-horizon 
journey. 

\subsection{Numerics}

We calculate the power fraction $S$ of isocurvature perturbations, and the 
cross-correlation $\cos\Delta$ between the adiabatic and the isocurvature 
perturbations at the end of the inflation epoch when the universe enters the 
radiation regime. These two quantities are defined by, respectively,
%eq53
\begin{equation}
S = \frac{\langle (\delta^{\rm en})^2\rangle}
{\langle (\delta^{\rm ad})^2\rangle +\langle (\delta^{\rm en})^2\rangle},
\end{equation}
%eq54
\begin{equation}
\cos\Delta =\frac{1}{\langle(\delta^{\rm ad})^{2}\rangle^{1/2}}
\frac{\langle \delta^{\rm en}\delta^{\rm ad}\rangle}
{\sqrt{\langle (\delta^{\rm ad})^2\rangle + 
   \langle(\delta^{\rm en})^2}\rangle }.
\end{equation}

Two sets of initial conditions, (42) - (44) and (48) - (50), are imposed 
respectively to calculate $S$ and $\cos \Delta$. Both quantities depend mainly 
upon the dissipation parameter $\Gamma_H$. The results are plotted in Figs. 1 
and 2. Figure 1 shows the prominent mixture and correlation of two distinguished 
perturbation components for the case where the fluctuations of the $\phi$ field 
and of the thermal bath are totally uncorrelated. The $\Gamma_H$-dependence of 
the curves for $S$ and $\cos\Delta$ in Fig. 2 reveals the similar behavior as 
in Fig. 1, but with different amounts. In particular, the isocurvature 
perturbations and the adiabatic ones are in phase when $\Gamma_H >0.217$, but 
are anti-correlated when $\Gamma_H < 0.217$. 

%fig1
\begin{figure}[ht]
\begin{center}
%\begin{turn}{-90}
\epsfig{file=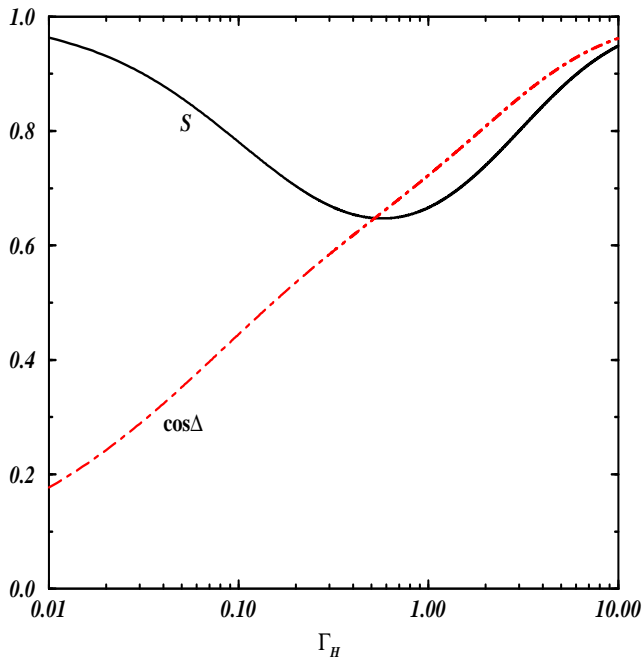,width=4in,height=4in}
%\end{turn}
\end{center}
%\leavevmode
%\hbox{
%\epsfxsize=6.0in
%\epsffile{fig1.eps}}
\caption{The power fraction $S$ of the isocurvature perturbations, 
and the cross-correlation $\cos\Delta$ between $\delta^{\rm ad}$ and 
$\delta^{\rm en}$ as a function of parameter $\Gamma_H$ at the end of 
thermal dissipative inflation for initial conditions (42) - (44).}
\label{fig1}
\end{figure}
%
%fig2
\begin{figure}
\begin{center}
%\begin{turn}{-90}
\epsfig{file=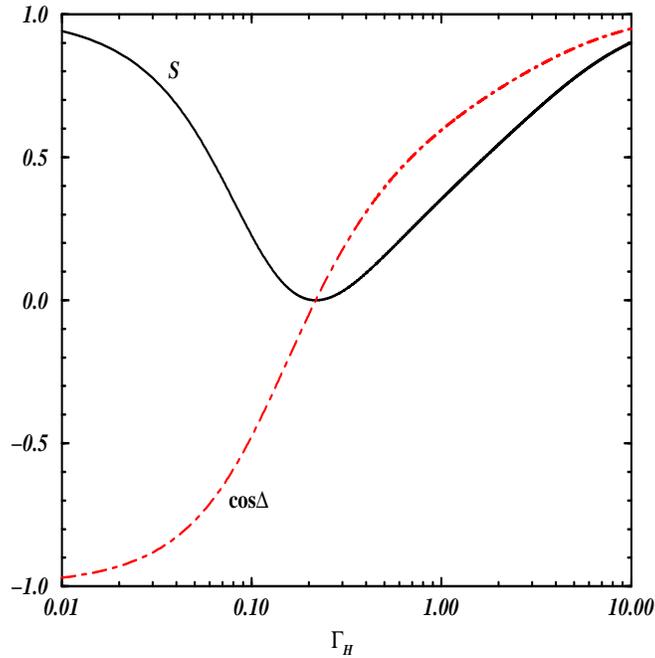,height=4in,width=4in}
%\end{turn}
\end{center}
%\leavevmode
%\hbox{
%\epsfxsize=6.0in
%\epsffile{fig2.eps}}
\caption{Same as Fig. 1 with initial conditions (48) - (50), i.e. taking 
into account the correlation given by the energy exchange between the 
background components. }
\label{fig2}
\end{figure}

\section{Conclusions and Discussions}

We show that the thermal dissipative inflation produces the initial hybrid perturbations with significant correlation. This is largely due to the 
coexistence of two components, the radiation and the $\phi$ field, during 
the inflationary epoch. The evolution of the density perturbation of radiation 
and that of the $\phi$ field on scales larger than horizon are governed by 
different equations (D3) and (D7). Consequently after reentering the horizon, 
the density perturbation of radiation and that originated from $\phi$ field 
generally are different. The super-horizon analysis is essential to reveral
the formation of the hybrid perturbations.

We have calculated the power fraction of the isocurvature perturbations 
($S$), and the correlation between the adiabatic and the isocurvature 
perturbations ($\cos\Delta$) at the end of the inflationary epoch. Since 
the transition from the inflationary era to the radiation-dominated period 
is smooth without an intervening reheating process, the values of 
$S$ and $\cos\Delta$ can be directly used as the initial conditions for the 
radiation regime. 

We found that, for each $k$-mode, $S$ and $\cos\Delta$ are mainly determined 
by the parameter $\Gamma_H$ at the instant as the perturbation mode crosses 
outside the Hubble horizon. That is, $S$ and $\cos \Delta$ depends entirely 
upon the dissipation of the inflaton. When taking into account the effect of 
energy exchange, $\cos \Delta$ is more sensitive to the dissipation parameter $\Gamma_H$. For strong dissipation cases where $\Gamma > 0.217H$ the 
adiabatic and the isocurvature perturbations are in phase. Under weak 
dissipation $\Gamma < 0.217H$, however, the two perturbation components are 
anti-correlated. 
 
Given that the current observational constraints on $S$ 
and $\cos \Delta$ are rather diverse\cite{pei,val,cro}, we do not make the 
detail parameter fitting in this paper, but disscuss some properties of the 
hybrid perturbations which are useful for further model testing. 
Apparently, in this thermal dissipative model $S$ and $\cos \Delta$ are not 
independent of each other. The predicted $S$-$\cos \Delta$ relations can 
directly be seen from Figs. 1 and 2. These relations hold for all $k$-modes, 
but they are not associated with the parameter $\Gamma$. On large scales
these predictions can be confronted with the CMB observation without
considering the effects of evolution of perturbation during the radiation
era. The relation $\cos \Delta \simeq \pm \sqrt{S}$ shown in Fig. 2 can 
then be tested by the observed adiabatic/isocurvature ratio and correlation. 
For instance, to realize the thermal dissipative inflation, the dissipation 
parameter $\Gamma_H$ can be taken in the range $\simeq 0.13-0.4$~\cite{bf,lf1,lf2}. 
Consequently, $S$ amounts to about 10\% and $\cos \Delta \simeq -0.3$ to +0.3. 

Secondly, the $k$-dependence of $\cos \Delta$ in this model is governed by 
the $k$-dependence of $\Gamma_H$. On the other hand, Eqs. (42)-(43) or 
(48)-(49) shows that the two differences 
\[
\frac{d\ln \langle (\delta^{\rm ad})^2\rangle}{d\ln k} - 
\frac{d\ln \langle (\delta^{\rm en})^2\rangle}{d\ln k} \hspace{1cm}{\rm and}\hspace{1cm} 
\frac{d^2\ln \langle (\delta^{\rm ad})^2\rangle}{d\ln k^2} -
\frac{d^2\ln \langle (\delta^{\rm en})^2\rangle}{d\ln k^2}   
\]
are also specified by the $k$-dependence of $\Gamma_H$. Therefore, the 
difference between the spectral indices, or the running spectral index, 
of the adiabatic and the isocurvature perturbations should also be 
determined by the $k$-dependence of the correlation $\cos \Delta$. For 
instance, if $\cos \Delta$ 
are $k$-independent for some range of $k$, the spectral indices should 
be fixed without any changes in that range. On the contrary, if there 
exists difference between the spectral indices of the two perturbation 
components, we should see a $k$-dependent $\cos\Delta$. This property is a 
robust prediction regardless of the initial conditions [Eqs. (42)-(44) 
or (48)-(50)] in use, and is effective to testing the thermal dissipative 
inflation model.   

\acknowledgments

WL is supported by the National Science Council, Taiwan, ROC under the 
Grant NSC91-2112-M-001-026.

\appendix

\section{Perturbation variables}

Appendixes A - D  will give all formulas needed for a full relativistic 
theory of linear perturbations for a system consisting of a $\phi$\ field 
and a fluid, such as radiation. Most material is taken from~\cite{b88,h123}.

\subsection{Perturbing the space-time metric}

The perturbed metric with respect to $\bar{g}_{ab}$ [Eq.(3)] is 
$g_{ab}=\bar{g}_{ab}+\delta g_{ab}$. In this paper, we are interested only 
in the scalar-type perturbations. For this purpose, the perturbations can 
be described by four variables\cite{b88}
%eqA1-4
\begin{eqnarray}
{\rm lapse \ function \ \alpha:} & g_{00}=-a^2(1+2\alpha) \\
{\rm  3-space \  curvature \ \varphi:} &
{\cal R}=\frac{4k^2}{a^2}\varphi \\ 
{\rm expansion \ scalar \  \kappa:} & \theta=3H-\kappa \\
{\rm  shear \ \chi:} & \ \ 
\sigma_{\alpha\beta}=\chi_{\left.\right|\alpha\beta}-\frac{1}{3}
g^{(3)}_{\alpha\beta}\chi^{\left.\right|\gamma}_{~\left.\right|\gamma} \\
  \nonumber
\end{eqnarray}
where the comoving spatial metric tensor $g^{(3)}_{\alpha\beta}$ is defined 
by $\bar{g}_{\alpha\beta}=a^2g^{(3)}_{\alpha\beta}$. The vertical bar
indicates a covariant derivative based on $g^{(3)}_{ab}$.

\subsection{Perturbing the scalar $\phi$ field}

 From Eq.(5), the linearly perturbed energy density is
%eqA5
\begin{equation}
\delta\rho_{(\phi)} =
  \frac{1}{2}\delta(\phi,_a u^a)^2 +V'\delta\phi.
\end{equation}
The first term on the r.h.s. of (A5) can be calculated by 
%eqA6
\begin{equation} 
\delta(\phi,_a u^a) =\delta\phi,_a u^a+ \phi,_a\delta u^a =
\delta\phi,_0u^0+\phi,_0 u^0\left(\frac{\delta u^0}{u^0} \right ).
\end{equation}
By means of (A1), we have $\delta u^0/u^0 \simeq 
-\delta g_{00}/ 2g_{00} = -\alpha$. Therefore, (A6) yields
%eqA7
\begin{equation}
\delta (\phi,_a u^a) = \delta\dot{\phi}-\dot{\phi}\alpha.
\end{equation}
Thus, (A5) becomes
%eqA8
\begin{equation}
\delta\rho_{(\phi)} =
 \dot{\phi}\delta\dot{\phi}-\dot{\phi}^2\alpha+V,_{\phi}\delta\phi.
\end{equation}

Similarly, the perturbation in the pressure of the $\phi$ field is 
%eqA9
\begin{equation}
\delta p_{(\phi)} =
\dot{\phi}\delta\dot{\phi}-\dot{\phi}^2\alpha-V,_{\phi}\delta\phi.
\end{equation}

For the background solution, $q_{(\psi) a}=0$. Since $q_{(\psi) a}u^a=0$, 
the linearly perturbed $q_{(\psi) a}=-\phi,_c u^c h^b_a\phi_{,b}$ can be 
described by
%eqA10
\begin{equation}
\delta q_{(\psi) \alpha} = \psi_{(\phi),\alpha},
\end{equation}
where
%eqA11
\begin{equation}
\psi_{(\phi)}=-\dot{\phi}\delta\phi.
\end{equation}
Since $ \pi_{(\phi)ab}=0 $ [Eq. (5)], the perturbed anisotropic pressure
$\delta \pi_{(\phi),ab}=0$.

\subsection{Perturbing the thermal bath}

 From Eq. (6), the perturbed variables of the thermal bath are given by
%eqA12-15
\begin{eqnarray}
{\rm perturbation \ of \  the \ energy \ density:} &  
 \varepsilon_{(r)}\equiv \delta\rho_{(r)}\  \\ 
{\rm perturbation \ of \ isotropic \ pressure:} & \
    \delta p_{(r)} = \frac{1}{3}\delta\rho_{(r)}= 
   \frac{1}{3} \varepsilon_{(r)}  \\
{\rm energy \ flux:}   & \delta q_{(r)a} =\psi_{(r),\alpha} \\ 
{\rm anisotropic \ pressure:} &  \ \ 
\delta \pi_{(r)ab}= 0. 
 \end{eqnarray}
where $\psi_{(r),\alpha}$ is the energy density flux of radiation. 

\subsection{Perturbing the interaction terms}

Using Eqs. (13) and (A6) the perturbed temporal component of the 
interaction term is
%eqA16
\begin{equation}
\delta Q_{(\phi)} = -\delta Q_{(r)}=- \Gamma \delta (\phi,_a u^a)^2=
 2\Gamma(\dot{\phi}\delta\dot{\phi}-\dot{\phi}^2\alpha).
\end{equation}
Similarly, the perturbed spatial component can be described by
%eqA17
\begin{equation}
\delta J_{(\phi)}= - \delta J_{(r)}= 
\Gamma \dot{\phi} \delta\phi.
\end{equation}

\section{Equations of perturbation variables}

Based on the perturbed variables given in \S A, the 
evolution of these variables is governed by the following equations:

\noindent {\it Definition of $\kappa$}
%eqB1
\begin{equation}
   \kappa=-3\dot{\varphi}+3H\alpha+\frac{k^2}{a^2}\chi,  
\end{equation}

\noindent {\it ADM energy constraint}
%eqB2
\begin{equation}
H\kappa-\frac{k^2}{a^2}\varphi=
 -4\pi G\left ( \varepsilon_{(r)} + \dot{\phi}\delta\dot{\phi}-
 \dot{\phi}^2\alpha+V,_{\phi}\delta\phi \right ),  
\end{equation}

\noindent {\it momentum constraint}  
%eqB3
\begin{equation}
\kappa-\frac{k^2}{a^2}\chi=
 12\pi G\left(\psi_{(r)}-\dot{\phi}\delta\phi\right),  
\end{equation}

\noindent {\it ADM propagation} 
%eqB4
\begin{equation}
\dot{\chi}+H\chi=\alpha+\varphi,  
\end{equation}

\noindent {\it Raychaudhuri equation} 
%eqB5
\begin{equation}
\dot{\kappa}+2H\kappa=\left(\frac{k^2}{a^2}-3\dot{H}\right)\alpha+
 4\pi G\left[2\varepsilon_{(r)}+4\dot{\phi}\delta\dot{\phi}
  -4\dot{\phi}^2\alpha-2V,_{\phi}\delta\phi \right],  
\end{equation}

\noindent {\it energy conservation of radiation} 
%B6
\begin{equation}
\dot{\varepsilon}_{(r)}+4 H\varepsilon_{(r)}=
  \frac{k^2}{a^2}\psi_{(r)}+(4/3)\rho_{(r)}(\kappa-3H\alpha)+
2\Gamma\dot{\phi}\delta\dot{\phi}-\Gamma\dot{\phi}^2\alpha, 
\end{equation}

\noindent {\it momentum conservation of radiation}
%B7
\begin{equation}
\dot{\psi}_{(r)}+3H\psi_{(r)}=
-(4/3)\rho_{(r)}\alpha-(1/3)\varepsilon_{(r)} 
   -\Gamma\dot{\phi}\delta\phi,
\end{equation}

\noindent {\it energy conservation of scalar field}
%B8
\begin{equation}
\delta\ddot{\phi}+(3H+\Gamma)\delta\dot{\phi}+\left(\frac{k^2}{a^2}
  +V''\right)\delta\phi=
  \dot{\phi}(\kappa+\dot{\alpha})-
  \left[(3H+\Gamma)\dot{\phi}+2V'\right]\alpha,
\end{equation}
The momentum conservation of scalar field yields an identity.  

\section{Gauge invariant variables}

Considering a gauge transformation $\widetilde{x}^a=x^a+\xi^a$, in which 
$\xi^a$ is an infinitesimal quantity, any tensor quantity ${\bf T}$
transforms as $\widetilde{{\bf T}} = {\bf T}-\cal L_{\xi}\it {\bf T}$,
where $\cal L_{\xi}\it {\bf T}$ is the Lie derivative of ${\bf T}$ in the
4-vector field $\xi^a$.  We may split the tensor ${\bf T}$
into its background and perturbed values; i.e.,
${\bf T} = \overline{{\bf T}}+\delta{\bf T}$. The transformation of
the perturbed parameter becomes
$\widetilde{\delta{\bf T}}=\delta{\bf T}-\cal L_{\xi}\it \overline{{\bf T}}$.
Applying the transformation to the metric tensor, we have
%eqC1
\begin{equation}
\widetilde{\delta g_{ab}}-\delta g_{ab} =
  -\cal L_{\xi}\it\bar{g}_{ab} \equiv -(\xi_{a;b}+\xi_{b;a})
= -(\bar{g}_{ac}\xi^c_{~,b}+\bar{g}_{cb}\xi^c_{~,a}+\bar{g}_{ab,c}\xi^c)~.
\end{equation}

Since the background 3-space is homogeneous and isotropic, all the perturbation
 variables are gauge independent under purely spatial gauge transformations. 
The perturbation equations presented above are also independent of the spatial 
gauge transformation. Under the temporal transformation factor 
$T_{\xi}\equiv a\xi^0$, the perturbed variables are given by 
~\cite{b88,h123}:
\begin{eqnarray}
%eqC2
\widetilde{\alpha}&=&\alpha-\dot{T}_{\xi}~, \\
%eqC3
\widetilde{\varphi}&=&\varphi-HT_{\xi}~, \\
%eqC4
\widetilde{\chi}&=&\chi-T_{\xi}~,  \\
%eqC5
\widetilde{\kappa}&=&\kappa+\left(3\dot{H}-\frac{k^2}{a^2}\right)T_{\xi}~, \\
%eqC6
\widetilde{\delta \rho}_i&=&\delta \rho_i-\dot{\rho}_{i}T_{\xi}~,  \\
%eqC7
\widetilde{\psi}_{i}&=&\psi_{i}+(\rho_{i} + p_{i})T_{\xi}~\\
%eqC8
\widetilde{\delta Q}_{i} & = & \delta Q_{i} - \dot{Q}_{i}T_{\xi}~\\
%eqC9
\widetilde{\delta J}_{i} & = &\delta J_{i} + Q_{i}T_{\xi}.
\end{eqnarray}
Equations (C6) to (C9) are applicable to components $(\phi)$, $(r)$ as well as 
the total fluids. 

Based on the time-gauge transformations [Eqs. (C2)-(C9)] and Background 
equations of motion [Eqs. (19) and (20)], one can construct the following 
gauge invariant variables: 
%eqC10
\begin{equation}
\zeta =
\varphi+ \frac{\delta\rho_{\rm tot}}{3(\rho_{\rm tot}+ p_{\rm tot})}=
\varphi+ \frac{\delta\rho_{(\phi)}+\delta\rho_{(r)}}
{3(\rho_{(\phi)}+p_{(\phi)}+ \rho_{(r)}+ p_{(r)})}
\end{equation}
%eqC11
\begin{equation}
\zeta_{(\phi)} =
\varphi+ \frac{\delta\rho_{(\phi)}+ 
 \delta J_{(\phi)}}{3(\rho_{(\phi)}+ p_{(\phi)})} 
\end{equation}
%eqC12
\begin{equation}
\zeta_{(r)} =
\varphi+ \frac{\delta\rho_{(r)}+\delta J_{(r)}}{3(\rho_{(r)}+ p_{(r)})}.
\end{equation}
%eqC13
\begin{equation}
\psi_{(\phi r)}= \frac{\psi_{(\phi)}}{\rho_{(\phi)}+p_{(\phi)}}
  -\frac{\psi_{(r)}}{\rho_{(r)}+p_{(r)}}
\end{equation}

\section{Equations of linear perturbation in the UCG}

Using the uniform-curvature gauge (UCG), the equations of the perturbed 
variables can be obtained from Eqs. (B1) to (B8) by setting up 
$\varphi\equiv 0$. We have
%eqD1
\begin{equation}
\alpha = \dot{\chi} + H\chi, 
\end{equation}
%eqD2
\begin{equation}
\kappa = 3H\dot{\chi} + \left(3H^2+\frac{k^2}{a^2}\right)\chi~,  
\end{equation}
%eqD3
\begin{equation}
\varepsilon_{(r)} = \left(\dot{\phi}^2-\frac{3H^2}{4\pi G}\right)\dot{\chi} +
  \left[\dot{\phi}^2-\frac{1}{4\pi G}\left(3H^2+\frac{k^2}{a^2}\right)
   \right]H\chi
-\dot{\phi}\delta\dot{\phi}-V'\delta\phi,  
\end{equation}
%eqD4
\begin{equation}
\psi_{(r)} = -\frac{H}{4\pi G}\dot{\chi}-\frac{H^2}
{4\pi G}\chi+\dot{\phi}\delta\phi, 
\end{equation}
%eqD5
\begin{equation}
\dot{\varepsilon}_{(r)}  = -4H\varepsilon_{(r)} + \frac{k^2}{a^2}\psi_{(r)}
 + \frac{4}{3}\rho_{(r)}\frac{k^2}{a^2}\chi + 2\Gamma\dot{\phi}\delta\dot{\phi} -
\Gamma\dot{\phi}^2\alpha,
\end{equation}
%eqD6
\begin{equation}
\dot{\psi}_{(r)} = -3H\psi_{(r)} - 
\frac{4}{3}\rho_{(r)}\left(\dot{\chi}+H\chi\right) -
\frac{1}{3}\varepsilon_{(r)} - \Gamma\dot{\phi}\delta\phi,
\end{equation}
%eqD7
\begin{eqnarray}
\delta\ddot{\phi} & = &
  -\left[(3H+\Gamma)-\frac{8\pi G}{3}
      \frac{\dot{\phi}^2}{H}\right]
  \delta\dot{\phi} - \left[\frac{k^2}{a^2}+V''+\frac{16\pi G}{3}
  \frac{\dot{\phi}V'}{H}\right]\delta\phi  \nonumber \\
& &-\left[\frac{32\pi G}{3} V\frac{\dot{\phi}}{H}+
  \left(\Gamma\dot{\phi}+2V'\right)\right]\dot{\chi} \nonumber \\
& &-\left[\left(\frac{32\pi G}{3} V- \frac{2}{3}\frac{k^2}{a^2}\right)\dot{\phi}+
  \left(\Gamma\dot{\phi}+2V'\right)H\right]\chi~,
\end{eqnarray}
%eqD8
\begin{eqnarray}
\ddot{\chi}&=&-\left[H+\frac{32\pi G}{3}\frac{V}{H}\right]\dot{\chi}-
  \left[4\pi G\left(\frac{8}{3}V-\dot{\phi}^2-\frac{4}{3}\rho_{(r)}\right)- 
\frac{2}{3}\frac{k^2}{a^2}\right]\chi
\nonumber \\
& & 
-4\pi G\left[-\frac{2}{3}\frac{\dot{\phi}}{H}\delta\dot{\phi}+
\frac{4}{3}\frac{V'}{H}\delta\phi\right],
\end{eqnarray}

 From (D1)-(D4), we have
%eqD9
\begin{equation}
\chi  =  \frac{4\pi G}{H}\frac{a^2}{k^2}
\left[3H(\psi_{(r)}+\psi_{(\phi)}) -(\varepsilon_{(r)}+\delta\rho_{(\phi)}) 
   \right ],
\end{equation}
and 
%eqD10
\begin{equation}
\dot{\chi} =  
-\frac{4\pi G}{H}\left (\psi_{(r)}+\psi_{(\phi)} \right )-H\chi.
\end{equation}


\begin{thebibliography}{99}

\bibitem{wmap} C. L. Bennett, {\em et al.}, astro-ph/0302207

\bibitem{dmr4} C. L. Bennett, {\em et al.} Astrophys. J. {\bf 464}, L1
(1996).

\bibitem{bri}  S.L. Bridle, A.M. Lewis, J. Weller \& G. Efstathiou,
   astro-ph/0302306

\bibitem{fj}  Y. P. Jing and L. Z. Fang, Phys. Rev. Lett. {\bf 73},
1882, (1994); L. Z. Fang and Y. P. Jing, Mod. Phys. Lett. {\bf A11},
1531 (1996).

\bibitem{bfh} A. Berera, L. Z. Fang and G. Hinshaw, Phys. Rev. {\bf D57},
2207 (1998).

\bibitem{pei} H.V. Peiris, {\em et al.}, astro-ph/0302225

\bibitem{kps} L. A. Kofman, Phys. Lett. {\bf B173}, 400(1986); D. Polarski 
  and A. A. Starobinsky, Nucl. Phys. {\bf B 385}, 623 (1992); D. Polarski
  and A. A. Starobinsky, Phys. Rev. {\bf D 50}, 6123 (1994). F. Di Marco, 
  F. Finelli, R. Brandenberger, Phys. Rev. {\bf D67}, 063512 (2003)  
  
\bibitem{double} D. Langlois, Phys. Rev. {D59}, 123512, (1999); D. Wands, 
 N. Bartolo, S. Matarrese \& A. Riotto, Phys. REv. {D66}, 043520, (2002);
 L. Amendola, C. Gordon, D. Wands \& M. Sasaki, Phys. Rev. Lett. {88}, 
 211302, (2002).

\bibitem{bf} A. Berera and L.Z. Fang, Phys. Rev. Lett. {\bf 74}, 1912 
(1995). A. Berera, Phys. Rev. Lett. {\bf D 75}, 3218 (1995).

\bibitem{lf1}  W. L. Lee and L. Z. Fang, Phys. Rev. {\bf D59}, 083503 (1999).

\bibitem{lf3}  W.L. Lee and L.Z. Fang, Class. Quant. Grav. 
   {\bf 17}, 4467, (2000)

\bibitem{hall} L.M. Hall, I.G. Moss and A. Berera, astro-ph/0305015

\bibitem{linde} J. Yokoyama \& A.D. Linde, Phys. Rev. {\bf D60},
    083509 (1999) 

\bibitem{ber3} A. Berera, Nucl. Phys. {\bf B585}, 666, (2000)

\bibitem{lf2} W.L. Lee and L.Z. Fang, Int. J. Mod. Phys. {\bf D6}, 305
  (1997)

\bibitem{tb} A.N. Taylor and A. Berera, Phys. Rev. {\bf D62}, 083517, (2000)

\bibitem{lee} W.L. Lee, thesis, Univ. of Arizona (1999).

\bibitem{el} G. F. R. Ellis and M. Bruni, Phys. Rev. {\bf D 40}, 1804
(1989); G. F. R. Ellis, M. Bruni \& J.-C. Hwang, Phys. Rev. {\bf D 42},
1035 (1990).

\bibitem{landau} L.D. Landau \& E.M. Lifshitz, {\em Statistical Physics}
  (Pergamon Press, 1989)

\bibitem{b88} J. M. Bardeen, in {\em Cosmology and Particle Physics},
ed. L. Z. Fang \& A. Zee, (Gordon and Breach Science Publishers, 1988)

\bibitem{h123} J.-C. Hwang, Astrophys. J. {\bf 375}, 443 (1991); J.-C.
Hwang, Astrophys. J. {\bf 415}, 486 (1993); J.-C. Hwang, Astrophys. J. {\bf
427}, 542 (1994).

\bibitem{val} see e.g. J. V\"{a}liviita \& V. Mubonen, astro-ph/0304175

\bibitem{cro} P. Crotty, J. Garc\'{i}a-Bellido, J. Lesgourgues, \& A. Riazuelo, astro-ph/0306286.

\end{thebibliography}
\end{document}